\documentstyle[emulateapj]{article}

\def\simlt{\lower.5ex\hbox{$\; \buildrel < \over \sim \;$}}
\def\simgt{\lower.5ex\hbox{$\; \buildrel > \over \sim \;$}}
\def\gsim{\;\rlap{\lower 2.5pt
\hbox{$\sim$}}\raise 1.5pt\hbox{$>$}\;}
\def\lsim{\;\rlap{\lower 2.5pt
   \hbox{$\sim$}}\raise 1.5pt\hbox{$<$}\;}

\def\spose#1{\hbox to 0pt{#1\hss}}
\def\lta{\mathrel{\spose{\lower 3pt\hbox{$\mathchar''218$}}
     \raise 2.0pt\hbox{$\mathchar''13C$}}}
\def\gta{\mathrel{\spose{\lower 3pt\hbox{$\mathchar''218$}}
     \raise 2.0pt\hbox{$\mathchar''13E$}}}
\newcommand{\beq}{\begin{equation}}
\newcommand{\eeq}{\end{equation}}


\lefthead{MENOU ET AL.}
\righthead{{``WEATHER'' VARIABILITY}}

\received{2002 August 30}
\begin{document}

\title{``Weather'' Variability Of Close-in Extrasolar Giant Planets}

\author{Kristen Menou\altaffilmark{a,1,2}, James
Y-K. Cho\altaffilmark{b,c}, Sara Seager\altaffilmark{c} \& Bradley M.S. Hansen\altaffilmark{d}}

\affil{$^{a}$Princeton University, Department of Astrophysical Sciences, Princeton, NJ 08544, USA}
\affil{$^{b}$Spectral Sciences Inc., 99 South Bedford
St., \# 7, Burlington, MA 01803, USA}
\affil{$^{c}$Carnegie Institution of Washington, Dept. of Terrestrial 
Magnetism, 5241 Broad Branch Rd. NW, Washington, DC 20015, USA}
\affil{$^{d}$Division of Astronomy, 8971 Math Sciences, UCLA, Los Angeles, CA 90095, USA}

\altaffiltext{1}{Chandra Fellow} 
\altaffiltext{2}{Present address: Department of Astronomy, P.O. Box
3818, University of Virginia, Charlottesville, VA 22903, USA}

\vspace{\baselineskip}

\begin{abstract}
Shallow-water numerical simulations show that the atmospheric
circulation of the close-in extrasolar giant planet (EGP) HD~209458~b
is characterized by moving circumpolar vortices and few bands/jets (in
contrast with $\sim 10$ bands/jets and absence of polar vortices on
cloud-top Jupiter and Saturn). The large spatial scales of moving
{circulation structures on HD~209458~b may generate detectable
variability of the planet's atmospheric signatures.} In this {\it
Letter}, we generalize these results to other close-in EGPs, by noting
that shallow-water dynamics is essentially specified by the values of
the Rossby ($R_o$) and Burger ($B_u$) dimensionless numbers. The range
of likely values of $R_o$ ($\sim 10^{-2}$--$10$) and $B_u$ ($\sim
1$--$200$) for the atmospheric flow of known close-in EGPs indicates
that their circulation should be {qualitatively similar to that of
HD~209458~b.}  This results mostly from the slow rotation of these
tidally-synchronized planets.

\end{abstract}

\keywords{planetary systems -- planets and satellites:
general -- stars: atmospheres -- turbulence}

\section{Introduction}

The focus of extrasolar planet research has broadened to now include
the characterization of their physical properties, as shown by the
recent {sodium detection in the atmosphere of} HD~209458~b
(Charbonneau et al. 2002). Atmospheric circulation is expected to play
a key role in determining a number of observational characteristics of
EGPs, including their albedo and transmission spectrum (see, e.g.,
Seager \& Sasselov 1998, 2000; Sudarsky et al. 2000; Brown 2001). This
is especially true for close-in EGPs, which are thought to be
tidally-locked to their parent star and irradiated on one side only:
circulation will be essential in redistributing heat from the day to
the night side on these planets, thus determining to a large extent
how they will appear to the distant observer (Cho et al. 2002a,b;
Showman \& Guillot 2002).

Recently, we have presented a set of detailed shallow-water numerical
simulations of the atmospheric flow on HD~209458~b (Cho et
al. 2002a,b), currently the only EGP with known mass and radius from
the transit light curves {and radial velocity measurements}
(Charbonneau et al. 2000; Henry et al. 2000; Mazeh et al. 2000; Jha et
al. 2000; Brown et al. 2001). These simulations suggest that, contrary
to the simple day/night (hot/cold) picture, the circulation on this
planet is characterized by two moving circumpolar vortices and a small
number of latitudinal bands/jets. The vortices act as dynamically
distinct thermal spots whose motion around the poles generates
variability as seen by an observer interested in quantities integrated
over the planetary disk (or circumference).

It is possible to determine the general features of the circulation
pattern expected within the framework of shallow-water dynamics by
specifying the two dimensionless numbers -- Rossby ($R_o$) and Burger
($B_u$) -- for the atmospheric flow. In this {\it Letter}, we estimate
a range of {likely $R_o$ and $B_u$ values} for known close-in EGPs and
conclude that their atmospheric circulation pattern {should be}
qualitatively similar to that of HD~209458~b. In \S2, we recall how
the atmospheric flow pattern can be characterized by the knowledge of
$R_o$ and $B_u$. In \S3, we describe the sample of close-in EGPs
selected for our study and how we estimate likely values for various
global planetary parameters entering into the definition of $R_o$ and
$B_u$. Finally, our results and conclusions are presented in \S4.

\section{Turbulent Shallow-Water Dynamics}

{Shallow-Water equations describe the motion of a thin, homogeneous
layer of hydrostatically-balanced, inviscid fluid with a free surface,
in motion around a rotating planet (Pedlosky 1987, Holton 1992).} The
fluid is subject to gravitational and Coriolis forces and obeys the
following equations
\begin{eqnarray}
\frac{\partial {\bf v}}{\partial t} + {\bf v \cdot \nabla v} & = & -g
{\bf \nabla} h - f {\bf k \times v},\\
\frac{\partial h}{\partial t}  +  {\bf v}\cdot\nabla\ h & = &
- h\nabla\cdot{\bf v},
\end{eqnarray}
where ${\bf v}$ is the horizontal velocity, $h$ is the thickness of
the modeled layer, $f=2 \Omega \sin \varphi$ is the Coriolis parameter,
$\Omega$ is the rotation rate of the planet, $\varphi$ is the latitude,
$g$ is the gravitational acceleration and ${\bf k}$ is the unit vector
normal to the surface of the planet. In dimensionless form,
shallow-water equations become functions only of the Rossby ($R_o$)
and Burger ($B_u$) numbers:
\begin{eqnarray}
R_o & \equiv & \frac{U}{|f| L},\\
B_u & \equiv & \left( \frac{L_D}{L} \right)^2,~L_D \equiv \sqrt{g H}/|f|,
\end{eqnarray}
where $U$, $L$ and $H$ are characteristic velocity, length and layer
thickness scales, respectively; $L_D$ is the Rossby deformation
radius. Note that $|f| \sim \Omega$ at mid-latitudes and that the
planetary radius, $R_p$, is the relevant {length scale when discussing
the large-scale  atmospheric circulation. The Rossby number
measures the importance of rotation on the flow, while the Burger
number measures the stratification of the atmosphere via the
Brunt-V\"ais\"al\"a frequency (Holton 1992).}

The atmospheric structure in bands of gaseous giant planets in our
Solar System is well described as emerging from freely-evolving
shallow-water turbulence on the sphere (Cho \& Polvani 1996b).
Turbulence in a thin atmospheric layer is quasi-2D in nature.
Contrary to the forward turbulent energy cascade observed in 3D
geometry, 2D turbulence is characterized by an inverse energy cascade
({transfer from small to} large scales) and a forward cascade of
enstrophy\footnote{The flow enstrophy is defined as $\frac{1}{2}
\xi^2$, where $\xi={\bf k \cdot \nabla \times v}$ is the flow
vorticity.} down to small scales, where it is dissipated by viscous
processes. {Qualitatively, the inverse cascade is associated with the
growth of vortices through continuous mergers.}

Turbulence in a thin atmospheric layer is also strongly constrained by
the combined effect of spherical geometry and rotation (the
``$\beta-$effect''). While the force balance is everywhere the same
along a latitude circle, it changes with latitude because of the
dependence of the Coriolis term with $\varphi$. This anisotropy, as
measured by the parameter $\beta=2 \Omega \cos \varphi / R_p$ (the
latitudinal gradient of $f$), is strongest {at the equator, where
$\beta$ is maximum.}. While fluid motions are free to grow to the
largest available scale in the longitudinal direction, their growth is
limited in the latitudinal direction by the characteristic Rhines
scale, $L_\beta=\pi \sqrt{2U/\beta}$ (Rhines 1975).  This {anisotropic
growth is a likely origin} of the banded structure on gaseous giant
planets in our Solar System; the number of bands expected for a given
planet is roughly $N_{\rm band} \sim \pi {R_p}/{L_\beta}$.

Cho \& Polvani (1996a) presented an extensive numerical study of
freely-evolving shallow-water turbulence. They explored the entire
parameter space of the equations, as determined by the two
dimensionless numbers $R_o$ and $B_u$. They showed that the anisotropy
due to the $\beta-$effect on a rotating sphere is necessary but not
sufficient to produce a long-lasting banded structure.  The Rossby
deformation radius, $L_D$, must also be $\lsim R_p/3$ for the banded
structure to be stable. This small value of $L_D$, which acts as a
{limiting scale for vortex interactions, prevents the formation (via
successive mergers) of large-scale structures such as circumpolar
vortices.}

Cho et al. (2002b) presented a generalization of the results of Cho \&
Polvani in the case when the planet is subject to day-side hemispheric
heating, as expected for close-in EGPs. Day-side heating was
prescribed in the adiabatic limit by forcing the fluid to be
permanently thicker on that side, while keeping the average thickness
constant. {Extensive exploration of the parameter space of the forced
model} showed that previous shallow-water dynamics results were
recovered even in the presence of this extra forcing (i.e.  $L_\beta$
and $L_D$ remain the relevant scales determining the number of bands
and the formation of polar vortices).

These results can be recast in terms of the values of $R_o$ and $B_u$
for the atmospheric flow. By setting $L \sim R_p$, $|f| \sim \Omega$
and $\beta \sim \Omega/R_p$ (mid-latitudes), we see that the number of
bands/jets expected is $N_{\rm band} \sim 1/\sqrt(2 R_o)$ and that the
presence of circumpolar vortices is expected for $B_u \gsim 1/9$.
Thus, if the values of $R_o$ and $B_u$ for other close-in EGPs can be
estimated, one can get an idea of the type of large-scale atmospheric
circulation expected on these planets in the stable, radiative
region. In our estimates of $R_o$ and $B_u$, we will set $U=\bar U$,
which is the global velocity scale of the atmospheric flow and is only
known for the Solar System giants (Table~\ref{tab:one}),

An important parameter entering the definition of both $R_o$ and $B_u$
is the planetary rotation rate, $\Omega$. While the value of $\Omega$
is generally unknown for EGPs, a number of close-in EGPs have the
advantage of being probably tidally-synchronized to their parent star,
so that their rotation rate has effectively been measured via the
orbital period ($\Omega=\Omega_{\rm orb}$ for circular orbits). As we
show below, the knowledge of $\Omega$ for this sample of close-in EGPs
restricts the range of possible values for $R_o$ and $B_u$ to a small
enough region of the parameter space that their atmospheric
circulation pattern can be {inferred}.

\section{Sample of Close-in Extrasolar Giant Planets}

Since tidal synchronization occurs faster than orbital
circularization, it is possible that some close-in EGPs with
substantial eccentricities are nonetheless (pseudo-)synchronized
(i.e. synchronized at the periastron orbital frequency).  The
shallow-water results described in \S2 were established only in the
limit of negligible eccentricity, however. Hence, we must restrict our
sample to planets with small eccentricities.\footnote{We limit the
eccentricity of planets in our sample to $e \lsim 0.1$, by analogy
with the small eccentricities of Solar System giants to which
``zero-eccentricity'' shallow-water models have been applied with
success (Cho \& Polvani 1996b; Cho et al. 2002b).}
Table~\ref{tab:one} lists all the EGPs selected for our study, plus
the four Solar System giants. Parameters for the EGPs were collected
from the extrasolar planet almanac\footnote{{\tt
http://exoplanets.org/almanacframe.html}} and
encyclopedia.\footnote{{\tt http://www.obspm.fr/encycl/encycl.html}}

Low-eccentricity EGPs were divided into two groups, based on {their
orbital distance to the parent} star. In the first group, EGPs with
semi-major axes $a \leq 0.066$~AU are most likely tidally-synchronized
since all known EGPs with such small values of $a$ are also
circularized. The tidal synchronization status of the more distant
planets (in the second group) is less clear because several eccentric
EGPs with distances of closest approach\footnote{The periastron
distance, where tidal forces are the strongest, is also roughly the
circular radius expected for the orbit after complete
circularization.} as small as $0.05$~AU are also known. It is thus not
clear why some EGPs with periastron distances larger than this value
would be tidally-circularized while others would not.  We note,
however, that for values of the tidal parameter $Q$ not too different
from that of Jupiter ($\sim 10^5$), EGPs in this second group are also
expected to be synchronized. We will assume it is indeed the case in
our calculations.

Radial velocity surveys only measure $M_p \sin i$, which is a lower
limit to the planet's mass, $M_p$, given the unknown orbital
inclination, $i$. For randomly oriented systems, the distribution of
{$\cos i$ is uniform. We adopt the value of $M_p$ corresponding to
$\sin i=0.5$ for our fiducial estimate of $R_o$ and $B_u$, and we
allow $\sin i$ to vary from $0.1$ to $1$} when estimating the range of
likely values for $R_o$ and $B_u$.\footnote{Note that for low values
of $\sin i$, some EGPs in Table~\ref{tab:one} have $M_p > 13 M_{\rm
Jup}$ and are thus brown dwarfs. We expect shallow-water results to be
applicable even in that limit.}

For a given mass, $M_p$, the radius of an isolated planet {is
estimated from the mass--radius relation for sub-stellar objects of
Chabrier \& Baraffe (2000), supplemented at the low mass end by a
constant density law that empirically fits values for the Solar System
giants.  To account for the slower cooling under strong stellar
irradiation, we also allow the radius to be up to $50\%$ larger than
the value for an isolated planet, in agreement with published cooling
EGP models (Burrows et al. 2000; Guillot \& Showman 2002). Our results
depend only weakly on the planetary radius, as long as $R_p \sim
R_{Jup}$ (as expected for all masses of interest). The} gravitational
acceleration is derived as $g=GM_p/R_p^2$, where $G$ is the
gravitational constant.

For the mean layer thickness, $H$, we adopt the atmospheric pressure
scale-height, $H_{\rm atm} \equiv {{\cal R} T_{\rm atm}}/{g}$, where
${\cal R}$ is the perfect gas constant. The global radiative
equilibrium temperature of the planet is $T_{\rm atm} = T_\star
({R_\star}/{2 a})^{1/2} (1-A_b)^{1/4}$, which is a function of the
parent star luminosity ($L_\star \propto T_\star^4 R_\star^2$), the
planet's semi-major axis $a$, and Bond albedo $A_b$. We adopt
$A_b=0.5$ for all our numerical estimates; our results only weakly
depend on the value of $A_b$ unless it approaches unity. The stellar
luminosity is derived from the mass through the simple mass-luminosity
relation $L_\star =(M_\star / M_\odot)^{3.6} L_\odot$.

The last two parameters needed to determine $R_o$ and $B_u$ are the
planetary rotation rate $\Omega$ and the global kinetic energy scale
$\bar U$. We assume that $\Omega =\Omega_{\rm orb} $ (as determined by
radial velocity surveys) in all cases. We allow $\bar U$ to vary from
$50$~m~s$^{-1}$, the smallest observed value for giant planets in the
Solar System (Jupiter), to $1000$~m~s$^{-1}$, a rather large value for
which the typical wind speeds in the atmosphere of hot, close-in EGPs
approaches the sound speed. A value $\bar U =400$~m~s$^{-1}$ is
adopted for our fiducial estimate of $R_o$ and $B_u$.

\section{Results}

\begin{figure}
\plotone{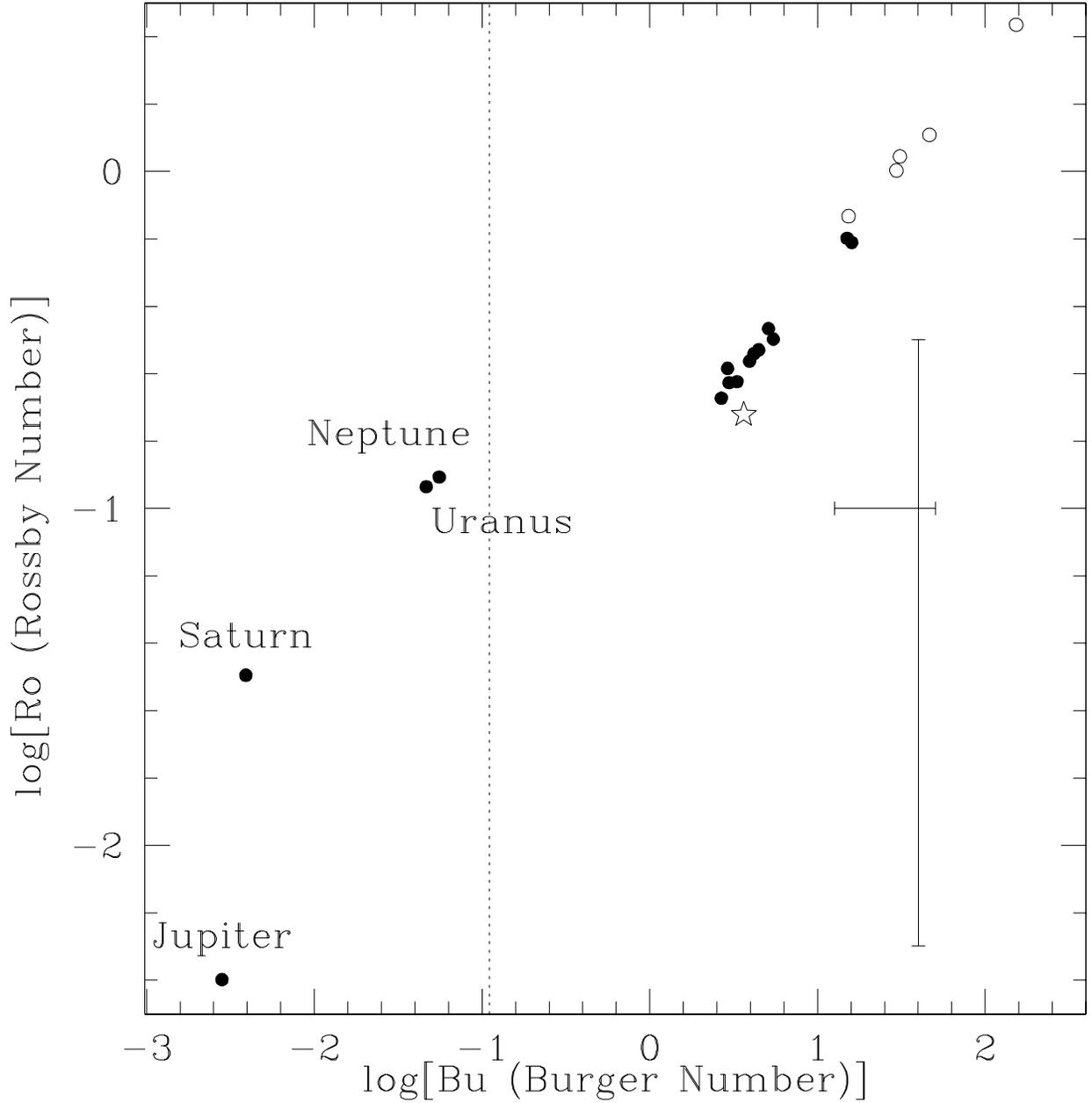}
\caption{Location of Solar System and close-in extrasolar giant
planets in the Rossby-Burger space. The assumption of tidal
synchronization for extrasolar planets represented by solid circles is
the safest (group 1; see Table~\ref{tab:one}). HD~209458~b is
indicated by a star. A representative range of possible values around
the fiducial estimates for extrasolar giant planets (each individual
solid or open circle) is shown as an errorbar (see Table~\ref{tab:one}
for details).  Formation of circumpolar vortices is expected in the
region to the right of the vertical dotted line ($B_u \gsim 1/9$). A
larger number of bands is expected for smaller values of $R_o$ (see
text for details).
\label{fig:one}}
\end{figure}

Estimated values for $R_o$ and $B_u$ are given in Table~\ref{tab:one}
for Solar System giants and close-in EGPs. The values listed for
close-in EGPs correspond to the range of min./max. values found given
the various assumptions detailed in \S3. Fiducial estimates are also
reported in figure~\ref{fig:one}, where solid dots correspond to group
1 EGPs (safe tidal synchronization assumption) and open circles to
group 2 EGPs (tidal synchronization assumption less safe). HD~209458~b
is indicated as a star.

It is clear from figure~\ref{fig:one} that close-in EGPs, as a group,
occupy a different region of the $R_o$--$B_u$ parameter space than
Solar System giants. In particular, they systematically have a Burger
number $B_u > 1/9$ (even when accounting for the large range of
allowed values; Table~\ref{tab:one}), which indicates that the
presence of circumpolar vortices is expected in the radiative region
{of close-in EGPs} within the framework of shallow-water dynamics. The
larger values of $R_o$ also indicate that generally few bands/jets are
expected on these planets (the uncertainty on $\bar U$ strongly
affects this number; see Table~\ref{tab:one}), thus allowing the
formation of larger ``great spots'' (which could also contribute to
the variability; Cho et al. 2002b). The near alignment of all the
points representing close-in EGPs in figure~\ref{fig:one} shows that
the dominant parameter determining their position in this diagram is
their rotation rate ($R_o \propto \Omega^{-1}$; $B_u \propto
\Omega^{-2}$). The small values of $R_o$ and $B_u$ for Solar System
giants reflect their relatively fast rotation rates.

Although we argued in favor of variable atmospheric signatures for
close-in EGPs, it is important to note that models do not yet
quantitatively predict how much variability is expected. In Cho et
al. (2002a,b), we emphasized that the combination of $\bar U$
(unknown) and the amplitude of day-night heating (parametrized in
adiabatic simulations) determines the contrast of the thermal spots
associated with circumpolar vortices. In the future, diabatic
shallow-water models will allow a self-consistent determination of the
day-night forcing. {More sophisticated models, combined with detailed
radiative transfer and chemistry descriptions (Seager \& Sasselov
1998; 2000; Seager et al. 2000), will allow us} to make quantitative
predictions regarding the level of variability expected for various
atmospheric signatures.

\section*{Acknowledgments}

Support for this work was provided by NASA through Chandra Fellowship
grant PF9-10006 awarded by the Smithsonian Astrophysical Observatory
for NASA under contract NAS8-39073.

\clearpage

\begin{table*}
{\tiny
\caption{GLOBAL PLANETARY PARAMETERS}
\begin{center}
\begin{tabular}{lcccccccccccc} \hline \hline
\\
Planet & $M_\star$ & $P_{\rm orb}$ & $a$ & $e$ & $M_p$ & $R_p$ & $g$ & $\Omega$
& $H$ & $\bar U$ &  $R_o$& $B_u$\\
 & ($M_\odot$) & (days) & (AU) & & ($M_{\rm Jup}$) & (m) & (m~s$^{-2}$) & (s~$^{-1}$) & (m) & (m~s$^{-1}$)& & \\
(1) & (2) & (3)& (4)& (5)& (6)& (7)& (8) & (9) & (10) & (11) & (12) & (13) \\
\\
\hline
\\
Neptune & 1.0 & 60,189.0& 30.05 & 0.0113& 0.054 & $2.5 \times 10^7$ & 11 & $9.75 \times 10^{-5}$ & $3 \times 10^4$ & 300 &$1.2 \times 10^{-1}$&$5.6 \times 10^{-2}$\\
Uranus & 1.0 & 30,685.4& 19.20 &0.0457& 0.046 & $2.6 \times 10^7$ & 9 & $(-)1 \times 10^{-4}$ & $3.5 \times 10^4$ & 300 &$1.2 \times 10^{-1}$&$4.8 \times 10^{-2}$\\
Saturn & 1.0 & 10,759.2& 9.58 &0.0565& 0.30 & $6.0 \times 10^7$ & 9 & $1.6 \times 10^{-4}$ & $4 \times 10^4$ & 300 &$3.2 \times 10^{-2}$&$3.9 \times 10^{-3}$\\
Jupiter & 1.0 & 4,332.6& 5.20 & 0.0489 & 1.00 & $7.1 \times 10^7$ & 23 & $1.8 \times 10^{-4}$ & $2 \times 10^4$ & 50 &$4.0 \times 10^{-3}$&$2.8 \times 10^{-3}$\\
\\                                             
\hline
\\
Group 2 &&&&&&&&&&&&\\
\\
\hline
\\
rho CrB b&0.95&39.6450&0.2300&0.0280&$>$1.100&--&--& $1.8 \times 10^{-6}$&--&--&0.24-7.4&68-180\\
HD 195019 b&1.02&18.3000&0.1400&0.0500&$>$3.430&--&--& $4.0 \times 10^{-6}$&--&--&0.10-4&30.8-69.6\\
Gl 86 b&0.79&15.7800&0.1100&0.0460&$>$4.000&--&--& $4.6 \times 10^{-6}$&--&--&0.09-3.4&13.2-46.4\\
55 Cnc b&0.95&14.6530&0.1150&0.0200&$>$0.840&--&--& $5.0 \times 10^{-6}$&--&--&0.084-2.8&13.2-39.2\\
HD 130322 b&0.79&10.7240&0.0880&0.0480&$>$1.080&--&--& $6.8 \times 10^{-6}$&--&--&0.062-2&6.8-18\\
\\                                             
\hline
\\
Group 1 &&&&&&&&&&&&\\
\\
\hline
\\
HD 168746 b&0.92&6.4090&0.0660&0.0000&$>$0.240&--&--& $1.1 \times 10^{-5}$&--&--&0.036-1.92&3.2-22.4\\
HD 49674 b&1.00&4.9400&0.0600&0.0000&$>$0.120&--&--& $1.5 \times 10^{-5}$&--&--&0.030-1.88&2.4-23.6\\
Ups And b&1.30&4.6170&0.0590&0.0340&$>$0.710&--&--& $1.6 \times 10^{-5}$&--&--&0.026-0.96&2.44-8.0\\
51 Peg b &0.95&4.2300&0.0512&0.0130&$>$0.440&--&--& $1.6 \times 10^{-5}$&--&--&0.024-1.04&1.64-7.6\\
HD 209458 b  & 1.05 & 3.5247 & 0.0450 & 0.0000 & 0.690 & $10^8$ & 8 & $2.1 \times 10^{-5}$ & $2 \times 10^6$ &--&0.024--0.48&3.6\\
HD 75289 b&1.05&3.5100&0.0460&0.0540&$>$0.420&--&--& $2.1 \times 10^{-5}$&--&--&0.024-0.88&1.32-6.0\\
BD -10316 b&1.10&3.4870&0.0460&0.0000&$>$0.480&--&--& $2.1 \times 10^{-5}$&--&--&0.020-0.84&1.36-6.0\\
Tau Boo b &1.30&3.3120&0.0500&0.0000&$>$4.090&--&--& $2.2 \times 10^{-5}$&--&--&0.019-0.72&1.36-4.96\\
HD 179949 b &1.24&3.0930&0.0450&0.0500&$>$0.840&--&--& $2.4 \times 10^{-5}$&--&--&0.018-0.62&1.20-3.56\\
HD 187123 b&1.06&3.0900&0.0420&0.0300&$>$0.520&--&--& $2.4 \times 10^{-5}$&--&--&0.019-0.72&1.08-4.4\\
HD 46375 b&1.00&3.0240&0.0410&0.0000&$>$0.249&--&--& $2.4 \times 10^{-5}$&--&--&0.017-0.90&1.0-6.4\\
HD 83443 b&0.79&2.9853&0.0380&0.0800&$>$0.350&--&--& $2.4 \times 10^{-5}$&--&--&0.017-0.80&0.80-6.4\\
\\
\hline
\end{tabular}
\label{tab:one}
\end{center}}
NOTES: (1) In order of decreasing orbital period (2) Stellar mass (3)
Orbital period (4) Semi-major axis (5) Eccentricity (6) Planet mass
(7) Planet radius (8) Surface gravity (9) Rotation rate (10)
Atmospheric scale-height (11) Global atmospheric velocity
scale (12) Rossby number (13) Burger number
\end{table*}
 
\clearpage
 
\end{document}